# Congestion Control in the Internet by Employing a Ratio-dependent Plant-Herbivore-Carnivorous Model

Shahram Jamali
Electrical and Computer Engineering Department
University of Mohaghegh Ardabili
Ardabil, Iran
jamali@iust.ac.ir

Morteza Analoui
Computer Engineering Department
Iran University of Science and Technology
Tehran, Iran
analoui@iust.ac.ir

*Abstract*—The demand for Internet-based services has exploded over the last decade. Many organizations use the Internet and particularly the World Wide Web as their primary medium for communication and business. This phenomenal growth has dramatically increased the performance requirements for the Internet. To have a high-performance Internet, a good congestion control system is essential for it. The current work proposes that the congestion control in the Internet can be inspired from the population control tactics of the nature. Toward this idea, each flow (W) in the network is viewed as a species whose population size is congestion window size of the flow. By this assumption, congestion control problem is redefined as population control of flow species. This paper defines a three-trophic food chain analogy in congestion control area, and gives a ratio-dependent model to control population size of W species within this plant-herbivore-carnivorous food chain. Simulation results show that this model achieves fair bandwidth allocation, high utilization and small queue size. It does not maintain any per-flow state in routers and have few computational loads per packet, which makes it scalable.

## I. INTRODUCTION

Bio-inspired computing is becoming an active area of research that focuses on the study of different self-organizing processes in nature, and using their principles as an inspirational metaphor to propose novel solutions to different daunting classical scientific problems. Initially limited to Particle Swarm Optimization (PSO) and Ant Colony Optimization (ACO), today Nature is serving as an inspiration source to a variety of other applications, ranging from computer science to complex logistic systems. The adapted mechanisms find applications in computer networking, for example, in the areas of network security [1, 2], pervasive computing, and sensor networks [3]. The central aim of this paper is to obtain methods for engineering congestion control algorithms, which have similar high stability and efficiency as biological entities often have.

Previous congestion control researches have been heavily based on measurements and simulations, which have intrinsic limitations [4]. There are also some theoretical frameworks and especially mathematical models that can greatly help us understand the advantages and shortcomings of current Internet technologies and guide us to design new protocols for identified problems and future networks [4, 5, 6, 7, 8, 9, 10, 11, and 12].

As another framework, a new class of congestion control algorithms was reported in [13, 14, 15, 16, 17 and 18], which is based on natural population control tactics. In the basic work of this research [13], we proposed a multidisciplinary conceptual framework that provides principles for designing and analyzing bio-inspired congestion control algorithms. We proposed that the natural population control tactics are susceptible to be mapped to the congestion control problem in the Internet. In [14, 15] inspiring by predator-prey interaction we developed a bio-inspired congestion control algorithm (BICC) and discussed on how a skillful parameters setting can help us to achieve good equilibrium properties such as fairness and performance. The limitation of BICC was its dynamic properties such as low smoothness and low speed of convergence. In [16] we have designed another bio-inspired congestion control algorithm that has better dynamic performance in compare with BICC. This algorithm that was called RBICC uses principals of a more *Realistic* model of predator-prey interaction. In other work [17], the intra-species and inter-species competition of *W* species was involved to complement BICC algorithm. The next branch of this research was started by paper [18], which employs a three-trophic food chain to design a congestion control scheme, called TTBICC (Tri-Trophic Biologically Inspired Congestion control).

The current paper is an improvement over paper [18]: while [18] uses a prey-dependent model of the three-trophic food-chain, this paper uses a ratio-dependent model. It has been concluded that natural systems are closer to the models with ratio-dependence than to the ones with prey-dependence. Hence, it is expected that the new congestion control algorithm, which we call it ETTBICC (Enhanced TTBICC), behaves better than TTBICC.

## II. PRELIMINARIES

In this section we first explain, how the Internet can be viewed as an ecosystem, and then we bring some natural population interaction models, which later will be used in design of our congestion control mechanism.

*A. Internet Ecosystem*

Consider a network with a set of *k* source nodes and a set of *k* destination nodes. We denote $S=\{S_1, S_2, ..., S_k\}$ as the set of source nodes with identical round-trip time (*RTT*) and $D=\{D_1, D_2, ..., D_k\}$ as the set of destination nodes. Our network model consists of a bottleneck link from LAN to WAN as shown in Figure 1 and uses a window-based algorithm for congestion control. The bottleneck link has capacity of *B* packet per *RTT*. The congestion window (*W*) is a sender-side limit on the amount of data the sender can transmit into the network before receiving an acknowledgment (*ACK*). All connections are assumed to be long-lived.

We imagine this network as an ecosystem that connects a wide variety of habitats such as routers, hosts, links and operating systems, etc. From congestion control viewpoint, there are some species in these habitats such as "Flow"(*W*), "Packet Drop"(*P*), "Queue"(*q*) and "Link Free Capacity" (*C*). The size of these network variables refers to their population size in the Internet ecosystem. In this ecosystem there is a whole web of interacting species and hence, their population sizes are affected. Figure 2 shows the typology of the Internet ecosystem from congestion control perspective.

Now, assume that population of *Flow* in source $S_i$ is $W_i$ (congestion window size of connection *i*). It is clear that if the population size of this species is increased, then the number of sent packet would be increased, too. Hence, in order to control the congestion in the communication networks the population size of $W_i$ (for *i*=1 to *k*) must be controlled. This means that *the population control problem in the nature can be mapped to the congestion control problem in the communication networks*. Nature uses many tactics such as predation, competition, etc. to control the population size of any species, which can be inspired to control the population size of *W* species. In papers [14, 15, 16 and 17] we have used predator-prey and competition interactions to control population size of *W* species. This work controls population size of *W* species within a three-trophic food chain and its preliminary version has been published in [18]. While paper [18] maps the prey-dependent food chain model to congestion control area, in this paper we use ratio-dependent food chain model for this purpose.

*B. Predator-Prey Interaction*

This interaction refers to classical predators that kill individuals and eat them:

**(1)** In the absence of predators, prey would grow exponentially. **(2)** The effect of predation is to reduce the prey's growth rate. **(3)** In the absence of prey, predators will decline exponentially. **(4)** The prey's contribution to the predator's growth rate is proportional to the available prey as well as to the size of the predator population. **(5)** Prey's carrying capacity puts a ceiling on the prey population.

If *r* and *f* represent the number of rabbits (prey) and foxes (predator), then this interaction can be explained by Lotka-Volterra model [19, 20]:

$$\frac{dr}{dt} = ar - brf \quad (1)$$

$$\frac{df}{dt} = crf - hf \quad (2)$$

Where *a* is the natural growth rate of rabbits, *b* is the death rate per encounter of rabbits due to predation, *c* is the efficiency of turning predated rabbits into foxes and *h* is the natural death rate of foxes in the absence of food. One of the unrealistic assumptions in the Lotka–Volterra model (1)-(2) is that the prey growth is unbounded in the absence of predation. As a reasonable step, we might expect the prey to satisfy a logistic growth [19, 20], say, in the absence of any predators, has some maximum carrying capacity:

$$\frac{dr}{dt} = a(1 - \frac{r}{C_r})r - brf \quad (3)$$

In which $C_r$ is the carrying capacity for the prey (*r*) when *f=0*.

*C. Three-Trophic Food Chain*

The classical food chain models with only two trophic levels are insufficient to produce realistic dynamics [21, 22]. Therefore we consider a three trophic levels food chain model. In an ecosystem there is always a foundation species of plant (*p*) that directly harvests energy from the sun, for example, grass. Next are herbivores (primary consumers) that eat the grass, such as the rabbits (*r*). Next are carnivores (secondary consumers) that eat the rabbits, such as a foxes (*f*). In this three-species food chain the lowest-level prey *p* is preyed upon by a mid-level species *r*, which, in turn, is preyed upon by a top-level predator *f*. The prey-dependent model of this ecosystem can be given as follows [23]:

$$\frac{dp}{dt} = p\left(\alpha\left(1 - \frac{p}{C_p}\right) - \beta r\right) \quad (4)$$

$$\frac{dr}{dt} = r\left(a\left(1 - \frac{r}{C_r}\right) + \varepsilon p - bf\right) \quad (5)$$

$$\frac{df}{dt} = f(cr - h) \quad (6)$$

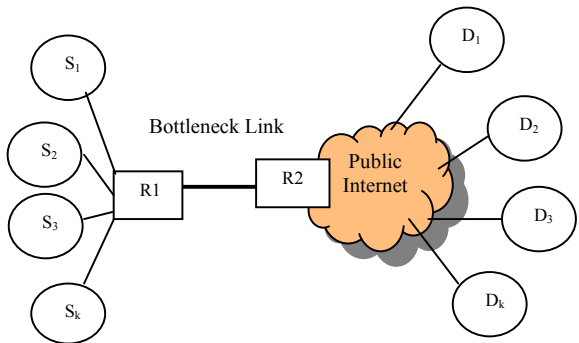

Figure. 1. Network Model

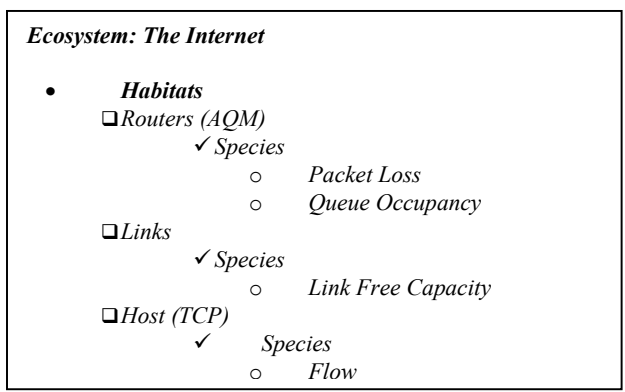

Figure 2. Internet Ecosystem Typology

In which $\alpha$ is the natural growth rate of *plant*, $\beta$ is the death rate per encounter of *plant* due to predation by rabbits, $C_p$ is the carrying capacity of *plant*, $\varepsilon$ is rate of turning predated *plants* into rabbits and *other* parameters are defined as before.

Experimental observations suggest that prey-dependent models are appropriate in homogeneous situations and ratio-dependent models are good in heterogeneous cases. By many investigators it has also been concluded that natural systems are closer to the models with ratio dependence than to the ones with prey dependence [24]. The ratio-dependent model of this ecosystem can be given in the following form [24]:

$$\frac{dp}{dt} = p\left(\alpha\left(1 - \frac{p}{C_p}\right) - \beta \frac{m_1 r}{p + n_1 r}\right) \quad (7)$$

$$\frac{dr}{dt} = r\left(a\left(1 - \frac{r}{C_r}\right) + \varepsilon \frac{m_1 p}{p + n_1 r} - b \frac{m_2 f}{f + n_2 r}\right) \quad (8)$$

$$\frac{df}{dt} = f\left(\frac{m_2 r}{r + n_2 f} - h\right) \quad (9)$$

For i=1,2, $m_i$ and $n_i$ are maximal predator growth rates and half-saturation constants.

### III. CONGESTION CONTROL MODEL

As mentioned, our previous work [18] uses a prey-dependent model of the three-trophic food chain to design the bio-inspired congestion control algorithm. Since the ratio-dependent model is more accurate than prey-dependent model, it seems that the next reasonable step is employing the ratio-dependent model of a plant-herbivore-carnivorous food chain to design a congestion control mechanism.

#### A. Plants-Herbivore-Carnivorous Analogy In The Internet Ecosystem

Current Internet congestion control mechanism consists of the congestion window algorithm of TCP running at end-system, and active queue management (AQM) algorithm at routers. Most of AQM algorithms (e.g. RED), use length of queue, at the congested router, as a congestion measure. Generally, interaction of TCP-AQM can be described as follows: **(1)** In the absence of waiting packets in the queue, congestion window (*W*) would grow. **(2)** When there is backlog of waiting packets in the queue, congestion window size would decline. **(3)** Incoming packet rate (congestion window size) contribution to the queue length growth is proportional to available traffic intensity, as well as, the queue length itself. **(4)** In the absence of packets stream (small congestion window size), the queue length will decline. **(5)** Bottleneck bandwidth is a limit for the packet rate. We see that this behavior is so similar to the predator-prey interaction.

Now we consider the congestion control mechanism from other viewpoint and discuss about interaction of "Link Free Capacity" and Flow. **(1)** In the absence of packets stream (small congestion window), free capacity of the link would grow. **(2)** When sending rate is increased (large congestion window), then free capacity of the link would decline. **(3)** Any growth of free capacity of link leads to packet rate increment at the sources. **(4)** In the absence of free capacity in the link, sending rate (congestion window size) will decline. **(5)** Bottleneck bandwidth is a limit for link free capacity (carrying capacity). We see that this behavior also is similar to the predator-prey interaction. Hence, "Link Free Capacity"-Flow-"Queue Length" is similar to a plant-herbivore-carnivorous food chain, in which, "Link Free Capacity" is preyed upon by Flow, which, in turn, is preyed upon by "Queue Length".

These similarities motivate us to use the tri-trophic food chain model to design congestion control scheme. For this purpose we assume that there are two species *"Virtual Capacity"(C)* and "Virtual *Queue"(Q)* that take part in tri-trophic food chain C-W-Q. In this ecosystem *C* contributes to increase of the population size of $W_i$ (*i*=1 to *k*), but Q contributes to decrease of population size of $W_i$(*i*=1 to *k*).

Paper [18] uses prey-dependent model of (4)-(6) and proposes its congestion control model in the form of Eqs. (10)-(12):

$$\frac{dC}{dt} = C\left(\alpha\left(1 - \frac{C}{Cc}\right) - \beta \sum_{j=1}^{k} W_j\right)\delta \quad (10)$$

$$\frac{dW_i}{dt} = W_i\left(a_i\left(1 - \frac{W_i}{C_W}\right) + \varepsilon C - bQ\right) \quad \text{where } i = 1,...,k \quad (11)$$

$$\frac{dQ}{dt} = Q\left(\sum_{i=1}^{k} c_i W_i - h\right)\delta \quad \text{Where } h \text{ Min}(B, Q + \Sigma W_i). \quad (12)$$

As an improvement over paper [18], we use the ratio-dependent model of (7)-(10), leading to congestion control model of (13)-(15),which we call it, ETTBICC (Enhanced TTBICC):

$$\frac{dC}{dt} = C\left(\alpha\left(1 - \frac{C}{Cc}\right) - \beta \sum_{j=1}^{k} \frac{W_j}{C + W_j}\right) \quad (13)$$

$$\frac{dW_i}{dt} = W_i\left(a_i\left(1 - \frac{W_i}{C_W}\right) + \varepsilon \frac{C}{C + W_i} - b \frac{Q}{Q + W_i}\right) \quad (14)$$

$$\frac{dQ}{dt} = Q\left(\varphi \frac{\sum_{i=1}^{k} c_i W_i}{Q + \sum_{i=1}^{k} c_i W_i} - \eta \frac{min(B, Q + \Sigma W_i)}{Q + min(B, Q + \Sigma W_i)}\right) \quad (15)$$

In which $\alpha$ is the growth rate of $C$, $\beta$ and $\delta$ are the decrement rate per encounter of $C$ due to $W$ and $Q$. $a_i$ is the growth rate of $W_i$, $\varepsilon$ is the increment rate per encounter of $W_i$ due to $C$ and $b$ is the decrement rate per encounter of $W_i$ due to $Q$. $c_i$ is the efficiency of turning predated $W_i$ into $Q$. $\theta$ and $\eta$ are death rate of $Q$ due to bottleneck capacity ($B$). $C_C$ and $C_W$ are the carrying capacities of $C$ and $W$, respectively. Note that, $Q$ and $C$ in (13)-(15) are not exactly queue length and unused capacity of network link. They are only two control variables, which are defined by inspiration from nature, to control and direct $W$ size to a stable point that satisfies characteristics such as fairness and performance.

In this model $C_W$ is a limit on the congestion window size. It is clear that the lower bound of $C_W$ is fair share of each source from bottleneck bandwidth size i.e. ($B/k$), and its upper bound is bottleneck bandwidth size ($B$). We assume that there is at least two competitor flows passing from bottleneck link, hence, $C_W = B/2$ seems to be a reasonable setting. We use the following settings for the parameters of proposed model:

$$\alpha = B, c_i = \delta = a_i = b = \varphi = \theta = \eta = 1, C_C = B$$
$$C_W = B/2 (optional), \varepsilon = \beta = 0.5 \quad (16)$$

By these settings, system of (13)-(15) has the following properties:

- Equations (13) and (16) say that $C$ can increase up to $B$ (bandwidth of bottleneck link), but this increment is offset by incoming packets rate $\sum W_i$.
- Equations (14) and (16) propose that $W_i$ population has multiplicative increase, but this growth will be inhibited by size of $Q$ and $W_i$ and will be accelerated by $C$.
- Equation (13) is similar to a formal definition of queue dynamics at a router.

The proposed model (13)-(15) is globally stable and converging to its equilibrium from any positive initial state. (Applying methods used in [24] can mathematically prove this convergence).

### B. Implementation Issues of ETTBICC

We now present a Source/AQM algorithm to implement the proposed model in the communication networks. Like TCP, ETTBICC is a window-based congestion control protocol. According to (14) source $i$ needs $C$, $Q$ and $C_W$ to update his congestion window ($W_i$) and these variables can only be maintained in the congested router, so each ETTBICC packet carries a congestion header, which is used to communicate congestion information from router to sources.

- **Hdr-C** = *The Last Value for C.*
- **Hdr-Q** = *The Last Value for Q.*
- **Hdr-BW** = *Congested Link Bandwidth.*

ETTBICC algorithm has three parts: router's algorithm, sender's algorithm and receiver's algorithm.

### Router's Algorithm:

The function of a ETTBICC router is to compute the control variables and feedback them to cause the system to converge to optimal efficiency and max-min fairness. ETTBICC router's algorithm is as follows:

At time RTT, 2RTT, 3RTT,... congested router:

1. Receives $W_i$ packets from all of the sources $s_i \in S$ that goes through bottleneck link (its capacity is $B$ packet per second).
2. Computes $C$ and $Q$.
3. Insert $C,Q$ and $B$ in header of all passing packets.

### Receiver's algorithm:

ETTBICC receiver is similar to a TCP receiver except that when acknowledging a packet, it copies the congestion header from the data packet to its acknowledgment. ETTBICC receiver:

1. Receives packets and extracts $B$, $C$ and $Q$ from their header.
2. Insert $B$, $C$ and $Q$ in sent ACK packets.

### Sender's Algorithm:

ETTBICC sender's function revolves around adjusting its sending rate based on received feedback. Sender:

1. Receives Ack packets and extracts $B$, $C$ and $Q$ from their header.
2. If there is any change in extracted values, re-computes new congestion windows Size:

$$\frac{dW_i}{dt} = W_i \left( a_i \left(1 - \frac{W_i}{C_W}\right) + \varepsilon \frac{C}{C+W_i} - b \frac{Q}{Q+W_i} \right) \text{ where } C_W = B/2$$

Note that, router's algorithm doesn't keep any per-flow state in the router and have few computational loads per packet, which makes it scalable.

## IV. ILLUSTRATIVE EXAMPLE

We now present preliminary simulation results to illustrate the promise of this approach. As paper [18], we apply ETTBICC to the network of Figure 1 and consider a four-connection network which has a single bottleneck link of capacity 50 pkts/RTT (for example if *RTT*=20 ms and packet size=1500 Byte, this capacity will refer to capacity of 30 Mbps). Other links of this network have bandwidth of 100 pkt/RTT. We suppose that all flows are long-lived, have the same end-to-end propagation delay and always are active. By using Eq. (16), congestion control system (13)-(15) can be adapted as follows for the test network:

$$\frac{dC}{dt} = C\left( \left(1 - \frac{C}{50}\right) - 0.5\left( \frac{W_1}{W_1+C} + \frac{W_2}{W_2+C} + \frac{W_3}{W_3+C} + \frac{W_4}{W_4+C} \right) \right) \quad (17)$$

$$\frac{dW_1}{dt} = W_1\left( (1 - W_1/25) + 0.5\frac{C}{C+W_1} - \frac{Q}{Q+W_1} \right) \quad (18)$$

$$\frac{dW_2}{dt} = W_2\left((1-W_2/25) + 0.5\frac{C}{C+W_2} - \frac{Q}{Q+W_2}\right) \quad (19)$$

$$\frac{dW_3}{dt} = W_3\left((1-W_3/25) + 0.5\frac{C}{C+W_3} - \frac{Q}{Q+W_3}\right) \quad (20)$$

$$\frac{dW_4}{dt} = W_4\left((1-W_4/25) + 0.5\frac{C}{C+W_4} - \frac{Q}{Q+W_4}\right) \quad (21)$$

$$\frac{dQ}{dt} = Q\left(\frac{W_1+W_2+W_3+W_4}{W_1+W_2+W_3+W_4+Q} - \frac{min(Q+W_1+W_2+W_3+W_4, 49)}{Q+min(Q+W_1+W_2+W_3+W_4, 49)}\right) \quad (22)$$

We use the following heterogeneous initial state, to solve this example in Matlab 7.1 environment:

$C=50, Q=50, W_1(0)=1, W_2(0)=2, W_3(0)=1, W_4(0)=3$

Note that in Eq. (22) we have considered 49 instead of 50. By this setting, 1/50 of bottleneck capacity has been left free to handle burst loads and also queued packets. On the other hand, by initializing of $Q=50$ the sources are forced to start slowly. After this slow start any source can adjust its sending rate based on received feedbacks (congestion notifications). The simulation results of this ETTBICC congestion control system are shown in Figs. (3) and (4). In order to reference to the results of these figures, we consider them from the following viewpoints:

**1. Fairness**: Evolution of congestion windows size is shown in Figure 3. It shows that $W_i$ (for i=1 to k) is converging to the equilibrium and in this equilibrium the bandwidth is shared equally among sources despite their heterogeneous initial state (max-min fairness [25, 26]).

**2. Utilization**: Figure 4 shows throughput and queue size evolution on the congested link. According to this figure, after the startup transient of the sources, utilization of bottleneck link remains always around 100%.

**3. Queue evolution**: As can be found in Figure 4 the queue size is zero in equilibrium and less than 3 packets in transient. This means that queuing delay and jitter is negligible. This figure shows that if we set the queue capacity of the congested router around 10 packets, then there won't be any packet drop in this scenario.

**4. Stability and Speed of Convergence:** As we can see in Figure 3 and Figure 4, the source rates, queue size and aggregate load on bottleneck bandwidth have decreasing oscillation level and track stable behavior. They are also extremely smoothed in the steady state. Some parts of the results of TTBICC that had been developed based on the equation (10)-(12) are shown in Figure 5 and Figure 6. Comparison of results shows that ETTBICC has better dynamic performance the TTBICC and converges fast to its fair and high-performance equilibrium.

V. CONCLUSION

In this paper we have designed a bio-inspired congestion control algorithm. Toward this design, the following steps were considered: (1) Redefinition of congestion control problem as population control of $W$ species. (2) Choosing a tri-trophic food chain as a natural population control framework to control population size of $W$ species. (3) Derivation a congestion control model from the ratio-dependent model of the tri-trophic food chain. (4) Analysis of equilibrium and dynamic properties of proposed model through a simulative study. Simulation results show that the proposed algorithm is fair and high-performance in its equilibrium. It also was observed that the proposed algorithm is globally converging to its equilibrium.

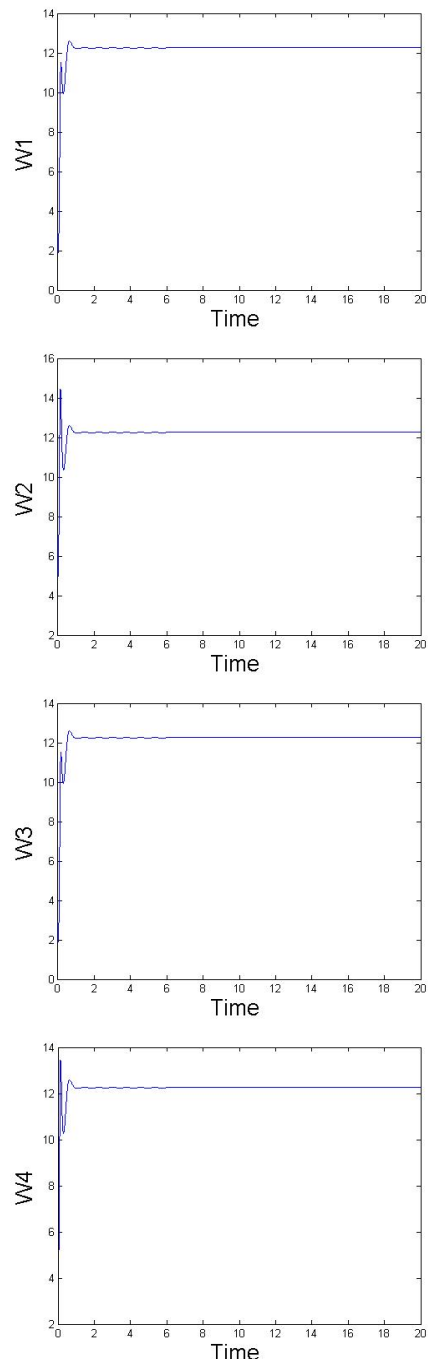

Figure 3. Evolution of ETTBICC's congestion windows size ($W_i$s)

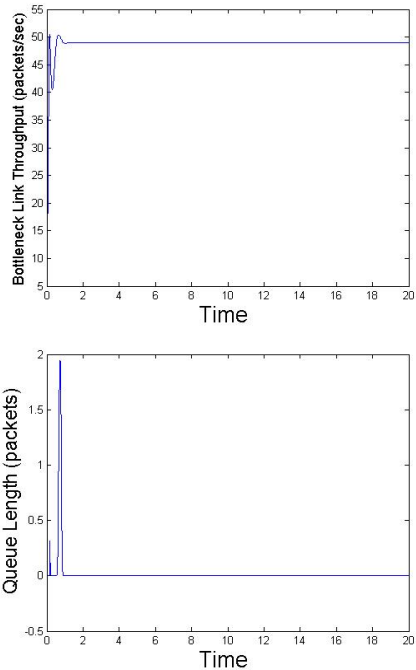

Figure 4. Throughput and queue length in ETTBICC's congested router

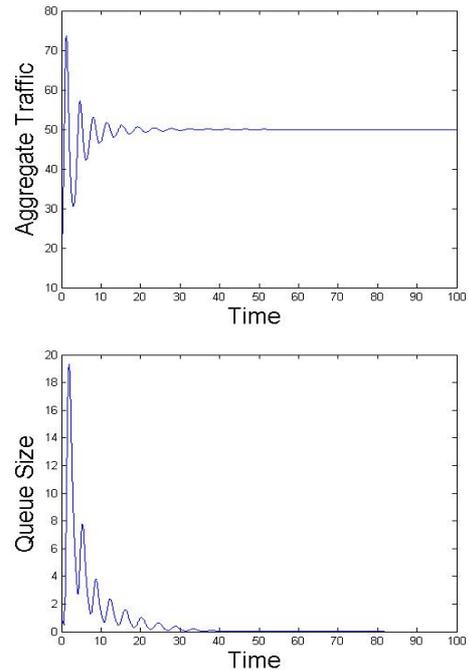

Figure 5. Evolution of TTBICC's throughput of bottleneck link and queue length [18]

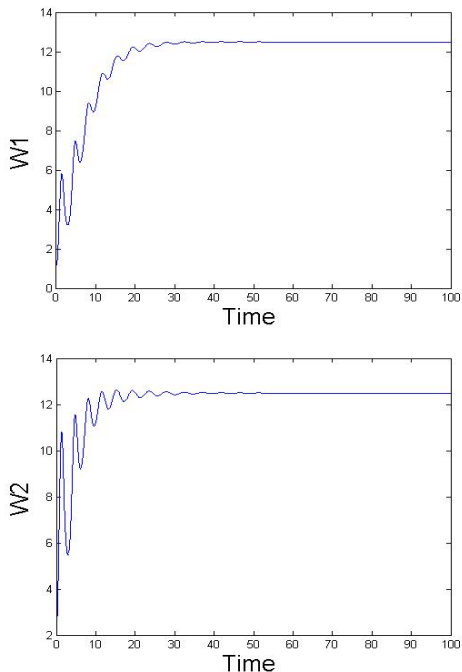

Figure 5. Evolution of TTBICC's congestion windows size ($W_i$s), [18]